\documentclass{rsproca_new}
\jname{rspa}
\usepackage{amsmath}
\usepackage{amssymb}
\usepackage{graphicx}
\usepackage{dcolumn}
\usepackage{bm}
\newcommand{\Sec}[1]{Section~\ref{sec:#1}}
\newcommand{\ssec}[1]{Subsection~\ref{sec:#1}}
\newcommand{\Fig}[1]{Figure~\ref{fig:#1}}

\newcommand{\Eq}[1]{Eq.\ (\ref{eq:#1})}
\newcommand{\Eqs}[2]{Eqs~(\ref{eq:#1})~and~(\ref{eq:#2})}

\newcommand{\Figs}[2]{Figures~\ref{fig:#1}--~\ref{fig:#2}}
\def\omegab{\mbox{\boldmath $\omega$}}
\def\Omegab{\mbox{\boldmath $\Omega$}}

\usepackage{latexsym}

\begin{document}

\title{Beyond Linear Fields: the Lie-Taylor Expansion}
\author{Wayne Arter}
\address{CCFE, Culham Science Centre, Abingdon, UK. OX14 3DB
}
\date{\today}

\begin{abstract}
The work extends the linear fields' solution of compressible
nonlinear magnetohydrodynamics~(MHD)
to the case where the magnetic field depends on superlinear
powers of position vector, usually but not always, expressed
in Cartesian components. Implications of the resulting Lie-Taylor series
expansion for physical
applicability of the Dolzhansky-Kirchhoff~(D-K) equations
are found to be positive. It is demonstrated how resistivity may be included
in the D-K model. Arguments are put forward that the D-K equations
may be regarded as illustrating properties of nonlinear MHD
in the same sense that the Lorenz equations inform about the
onset of convective turbulence. It is suggested that the Lie-Taylor
series approach may lead to valuable insights into other fluid models.
\end{abstract}

\keywords{ideal MHD, nonlinear, analytic solution, catastrophe, resistivity}

\corres{Wayne Arter}
\email{wayne.arter@ukaea.uk}

\maketitle

\section{Introduction}\label{sec:intro}
The recent work~\cite{Wa13a} showed how the equations of
ideal, compressible magnetohydrodynamics may be elegantly formulated
in terms of Lie derivatives, building on the work of Helmholtz, Walen
and Arnold. 
The magnetic induction equation for compressible flow may
be formulated in terms of
a Lie derivative of a vector by introducing the field~$\tilde{\bf B}$
defined as the magnetic field~${\bf B}$ divided by the mass density,
\begin{equation}\label{eq:induc}
\frac{\partial\tilde{\bf B}}{\partial t}=\mathcal{L}_{\bf u}(\tilde{\bf B})
\end{equation}
where $\mathcal{L}_{\bf u}$ is the Lie derivative with respect to the
flow field~${\bf u}$, $\tilde{\bf B}={\bf B}/\rho$ and $\rho$ is mass density.
The dynamical, potential vorticity equation~\cite{Wa13a}
may also be put into the Lie derivative form
\begin{equation}\label{eq:fullpv}
\frac{\partial \tilde{\boldsymbol{\omega}}}{\partial t}=\mathcal{R}+{\mathcal{L}}_{\bf u}(\tilde{\boldsymbol{\omega}})-{\mathcal{L}}_{\tilde{\bf B}}(\tilde{\bf J})
\end{equation}
where the potential vorticity $\tilde{\boldsymbol{\omega}}=\nabla\times{\bf u}/\rho$
and the potential current $\tilde{\bf J}=\nabla\times{\bf B}/\rho$. The
term~$\mathcal{R}$ vanishes either upon making the barotropic assumption
that pressure~$p(\rho)$ or sometimes in the isentropic approximation. 
Observe that the vectors which are evolved by \Eqs{induc}{fullpv} satisfy
$\nabla\cdot(\rho{\bf F})={\bf 0}$, provided there is mass conservation and
$\rho{\bf F}$ is solenoidal initially.

Now it is known since Dungey~\cite{dungey} that if the velocity field depends linearly
on Cartesian position vector, then compressible MHD is reducible
exactly to a set of ordinary differential equations~(ODEs) in the
coefficients of the proportionality constants.
(There is a much longer and complicated history
regarding classical hydrodynamics which will not be discussed herein.)
This ``linear fields" theory has been developed further as
described in Arnold \& Khesin~\cite[\S\,I.10.C]{arnoldkhesin} to 3-D.
Dolzhansky~\cite{Do05mech} explains clearly how a special choice of 3-D 
vector basis for both velocity and magnetic field
leads to the Kirchhoff equations,  a sixth order ODE describing the motion
of an ellipsoid immersed in fluid. The difficulty with the 3-D vector
basis is that it requires initial current
distributions corresponding to the linear magnetic field that are
not easy to realise in practice. It is possible to conceive that
an ellipsoidal blob of uniform vorticity might somehow appear,
indeed an \emph{elliptical} blob, embedded in a 2-D potential flow was proposed
by Helmholtz in~1889 as a model for a tornado~\cite[\S\,159]{lamb}.
However, it strains the imagination as to how an isolated \emph{ellipsoidal} current distribution
might be spontaneously produced. 

Arter~\cite{Wa87d} pointed out that the Cartesian linear fields could be regarded as
the truncation of a Taylor series expansion solution in position to first order.
Thus a problem with boundary conditions at a finite distance from the
origin might be formulated, by allowing higher order Taylor terms
to help say fix the current on a flat surface at distance~$\lambda$ rather
than on the problematic ellipsoidal surface.
Substituting the higher order terms in the governing partial
differential equations~(PDEs) leads to 
complicated sets of ordinary differential equations~(ODEs), but fortunately
the use of the above Lie derivative form for MHD is
especially convenient for such analysis, see the next \Sec{deriv}.
\ssec{case} first discusses implications of the results in \ssec{finite} of \Sec{deriv} for Dolzhansky's model
and then explores important mathematical features of the Dolzhansky-Kirchhoff~(D-K) equations.
\Sec{resist} discusses the introduction of resistivity into the D-K model
and conclusions are drawn in \Sec{conc}.

\section{Lie-Taylor Expansion and Implications}\label{sec:deriv}
\subsection{Lie Derivative Expansions}\label{sec:liederiv}
Suppose that the vectors ${\bf u}$ and ${\bf q}$ have components labelled~$j$
and moreover that each component may be separately expressed as 
a Taylor series in coordinates~$x^i$, i.e.
\begin{equation}\label{eq:utay}
u^j=U^j+U^j_{m}x^m+U^j_{mn}x^mx^n+U^j_{mnq}x^mx^nx^q+\ldots
\end{equation}
\begin{equation}\label{eq:qtay}
q^j=Q^j+Q^j_{m}x^m+Q^j_{mn}x^mx^n+Q^j_{mnq}x^mx^nx^q+\ldots
\end{equation}
using the Einstein summation convention, and where for example
\begin{equation}\label{eq:cptdefn}
U^j_{mn}= \frac{1}{2!}\frac{\partial^2 u^j}{\partial x^m \partial x^n}
\end{equation}
so that the suffices denote normalised derivatives with respect to
position coordinate. (More conventional notation would see a comma preceding
the suffices and no factorial prefactor, but here would serve to make
complicated expressions even longer.)
When ${\bf q}$ is set equal to~$\tilde{\bf B}$, the
evolution of ${\bf q}$ from \Eq{induc} may be written
\begin{equation}
\frac{\partial {\bf q}}{\partial t}=[{\bf u},{\bf q}]
\end{equation}
since there is the textbook result, see ref~\cite{Wa13a}, that
\begin{equation}\label{eq:lieder}
{\mathcal{L}}_{\bf v}({\bf w})^i= w^k\frac{\partial v^i}{\partial x^k}-v^k\frac{\partial w^i}{\partial x^k}
\end{equation}
which serves also to define the commutator~$[{\bf v},{\bf w}]$ for 
general vectors. Hence in component form
\begin{equation}\label{eq:comp}
\frac{\partial q^j}{\partial t}=
q^i\frac{\partial u^j}{\partial x^i}-u^i\frac{\partial q^j}{\partial x^i}
\end{equation}

It follows that it is necessary to calculate the derivative Taylor series:
\begin{eqnarray}\label{eq:diff1beg}
\frac{\partial q^j}{\partial x^k}&=&
Q^j_{m}\delta_{mk}+Q^j_{mn}\delta_{mk}x^n+Q^j_{mn}x^m\delta_{rnmk}+\nonumber \\
&+&Q^j_{mnq}\delta_{mk}x^nx^q+Q^j_{mnq}x^m\delta_{nk}x^q+Q^j_{mnq}x^mx^n\delta_{qk}+\ldots
\end{eqnarray}
which since partial derivatives commute, implying $Q^j_{kn}=Q^j_{nk}$, gives
\begin{equation}\label{eq:diff1}
\frac{\partial q^j}{\partial x^k}=Q^j_k+2Q^j_{kn}x^n+3Q^j_{knq}x^nx^q+\ldots
\end{equation}

If the Taylor series representations \Eq{qtay}, \Eq{diff1} and equivalents,
are now substituted in \Eq{comp}, the term which is independent
of position vector gives ODEs
\begin{equation}\label{eq:order0}
\dot{Q}^j=Q^iU^j_i -U^iQ^j_i
\end{equation}
Note that no truncation is needed i.e. that \Eq{order0} is exact,
regardless of the order of Taylor series truncation, but unless
$U^i=0$ (implying a flow stagnation point), $Q^j$ depends on~$Q^j_i$, for which an equation is needed.
This new equation may be obtained as the next step
in a procedure which forms successive positional
derivatives of the Taylor series representations \Eq{utay} and \Eq{qtay}
which are then substituted in successive positional
derivatives of \Eq{comp}, and at each order equates the constant terms.

Hence is formed first
\begin{equation}\label{eq:diff2}
\frac{\partial^2 q^j}{\partial x^l \partial x^k}= 2Q^j_{kl}+6 Q^j_{kln}x^n+\ldots
\end{equation}
which is then substituted together with \Eq{qtay}, \Eq{diff1} and equivalents in
\begin{equation}\label{eq:order1sta}
\frac{\partial }{\partial x^k} \frac{\partial q^j}{\partial t}=
\left[\frac{\partial u}{\partial x^k}, q\right]^j + \left[u, \frac{\partial q}{\partial x^k}\right]^j
\end{equation}
giving equations
\begin{equation}\label{eq:order1}
\dot{Q}^j_k=  Q^i_kU^j_i -U^i_kQ^j_i + 2( Q^iU^j_{ki}- U^i Q^j_{ki} )
\end{equation}
since for example
\begin{equation}\label{eq:order1exp}
\left[\frac{\partial u}{\partial x^k}, q\right]^j =
\frac{\partial u_i}{\partial x^k} \frac{\partial  q^j}{\partial x^i}-
q_i \frac{\partial^2 u^j}{\partial x^i\partial x^k}
\end{equation}
At next order
\begin{equation}\label{eq:diff3}
\frac{\partial^3 q^j}{\partial x^s \partial x^l \partial x^k}= 6 Q^j_{kls}+\ldots
\end{equation}
hence after the indicated manipulations and upon division by~$2$,
\begin{equation}\label{eq:order2}
\dot{Q}^j_{kl}=
 Q^i_lU^j_{ki} -U^i_kQ^j_{il} + Q^i_kU^j_{il} -U^i_{l}Q^j_{ik} 
 +Q^j_iU^i_{kl} -U^j_iQ^i_{kl} + 3(Q^iU^j_{ikl} -U^i Q^j_{ikl})
\end{equation}
and so on.

At $N^{th}$ order, it is apparent that 
coefficients with $N$ suffices evolve according to sums of nonlinear terms each 
containing a total of $N+1$ suffices.
It follows that if $U^i=Q^i=0$, then there is no closure problem, each order
varies in time depending only on itself and lower order contributions.

Equations for the evolution of $Q^j Q^j$, $Q^l_i Q^j_k$ (summation convention)
are also of interest, viz.
\begin{equation}\label{eq:ordsq0}
Q^j\dot{Q}^j=Q^iQ^jU^j_i -U^iQ^jQ^j_i 
\end{equation}
and in the case where $U^i=Q^i=0$,
\begin{eqnarray}\label{eq:ordsq1}
Q^l_i\dot{Q}^j_k+\dot{Q}^l_iQ^j_k &=&
Q^l_iU^j_iQ^i_k
-Q^l_i Q^j_iU^i_k\nonumber\\
 &+& U^l_iQ^i_iQ^j_k -Q^l_i U^i_iQ^j_k
\end{eqnarray}
Strategically relabelling $i$ and~$j$ shows that the terms in $QUQ$ cancel,
and hence
\begin{equation}\label{eq:commsq}
d{Q^2}/dt=\dot{Q^2}=[U,Q^2]
\end{equation}
where $[.,.]$ is the $3\times3$ matrix commutator. This result could have
been deduced directly from \Eq{order1} as a commutator equation 
\begin{equation}\label{eq:commutator}
\dot{Q}=[U,Q]
\end{equation}
whence it follows that for all integer~$L>0$
\begin{equation}\label{eq:commpower}
d{Q^L}/dt=\dot{Q^L}=[U,Q^L]
\end{equation}
Since the trace of the matrix commutator vanishes,
the trace of $Q^2$ (as well as $Q$) is constant in time (provided $U^i=Q^i=0$).
The more powerful result would
be that $tr(QQ^T)$ is conserved, instead
$tr(Q^2)$ is the sum of squares of 
elements of~$S$ less the sum of squares of elements of~$A$, where $S$ and $A$
are the symmetric and skew-symmetric parts of~$Q$ respectively.
Considering the separate cases
$Q=S$ and $Q=A$ together with need for the time evolution to maintain the (skew-)symmetry,
\Eq{commsq} implies that solutions of \Eq{order1} are bounded 
if $Q$ and $U$ are both skew-symmetric. When they are both symmetric,
their commutator is skew-symmetric and there is no consistent dynamic.
A stronger result can be deduced by contracting \Eq{order1} with~$Q^j_k$, giving
\begin{equation}\label{eq:order1sq}
Q^j_k \dot{Q}^j_k=  Q^j_k Q^i_kU^j_i -Q^j_kU^i_kQ^j_i
\end{equation}
As $Q^j_k Q^i_k$ and $Q^j_kQ^j_i$ are both symmetric tensors,
the sum of the squares of the matrix elements is conserved provided $U$
is skew-symmetric.

Since they are based purely on Taylor series' manipulations,
and the Lie derivative has the same form in any nondegenerate coordinate system
if the vector components are `raised', i.e. treated as contravariant,
it should be clear that all the above results apply in an arbitrary coordinate system.
The induction equation may also~\cite{Wa13c} be expressed as the vanishing of a 4-D Lie derivative,
with implications for deriving solutions which are polynomial in time.

\subsection{Scalar Transport Equation}\label{sec:scalar}
Although not strictly needed in the current work, the compressible MHD equations
are completed by scalar transport equations for quantities such as internal energy,
and at minimum the mass density~$\rho$. For consistency with the development
in the previous section, it is necessary to introduce the point mass
$\varrho=\sqrt{g} \rho$, for then (provided the volume element~$\sqrt{g}$ is independent
of time) $\varrho$ evolves as
\begin{equation}\label{eq:mtran}
\frac{\partial \varrho}{\partial t}+
\frac{\partial (\varrho u^i)}{\partial x^i}=0
\end{equation}
which is again an equation true in any reasonable coordinate frame. It is
worth remarking that the solenoidal constraint on $\rho{\bf F}$ is also frame
independent in this same sense, becoming 
\begin{equation}\label{eq:fdiv}
\frac{\partial (\varrho F^i)}{\partial x^i}=0
\end{equation}
so that all the vector fields in the Lie formulation of MHD satisfy \Eq{fdiv} in steady state.

Taylor expanding the point mass
\begin{equation}\label{eq:mtay}
\varrho=\varrho_0+\varrho_{m}x^m+\varrho_{mn}x^mx^n+\varrho_{mnq}x^mx^nx^q+\ldots
\end{equation}
substituting in \Eq{mtran} and equating coefficients as before, gives the hierarchy
\begin{eqnarray}\label{eq:mhier}
\dot{\varrho}_0&=&-\varrho_{0}U^j_j-\varrho_j U^j \nonumber\\
\dot{\varrho}_k&=&-2\varrho_{0}U^j_{jk}-\varrho_k U^j_j- \varrho_j U^j_k -2 \varrho^j_k U^j\\
\dot{\varrho}_{kl}&=&-3\varrho_{0}U^j_{jkl}-\varrho_{l} U^j_{jk}- \varrho_k U^j_{jl}
- \varrho_{kl} U^j_i - 3 \varrho^j_{kl} U^j \nonumber\\
\ldots &=& \ldots
\end{eqnarray}
Note again that for a consistent truncation it is necessary that $U^j=0$, or for all higher
derivatives of~$\varrho$ to vanish, and for higher order derivatives of~$U$ also
to be zero.

In the cases considered, $\sqrt{g}$ is both time and position independent so that
\Eq{mhier} applies with density $\rho$ replacing~$\varrho$. Moreover, the
model mainly considered is inherently incompressible, so $\rho$ is also
time and position independent.

\subsection{Vorticity and Current}\label{sec:VandC}
In a general non-orthogonal coordinate system with
coordinates~$x^j$, the curl
operator relating velocity to vorticity, is such that
\begin{equation}\label{eq:curlgg}
\omega^i=\frac{1}{\sqrt{g}} e^{ijk} \frac{\partial(g_{k l}u^l)}{\partial x^j}
\end{equation}
where $e^{ijk}=e_{ijk}$ is the permutation symbol ($e_{123}=-e_{132}=1$, $e_{112}=0$ etc.),
$g_{ij}$ is the metric tensor and $\sqrt{g}$ is the volume element.
Introducing the tensor $G_{ij}=g_{ij}/\sqrt{g}$, \Eq{curlgg} becomes after relabelling
\begin{equation}\label{eq:curlggx}
\omega^j=e^{jrs} G_{sq} \frac{\partial u^q}{\partial x^r}
+e^{jrs} u^q \frac{\partial G_{sq}}{\partial x^r}
+e^{jrs} G_{sq} u^q \frac{\partial(\ln\sqrt{g})}{\partial x^j}
\end{equation}

Supposing $g_{sq}$ to be constant, equivalently assuming an
affine transformation, the second two terms vanish, then if the linear fields'
assumption is made for the velocity
\begin{equation}\label{eq:linvort}
\omega^j=e^{jrs} G_{sq} U_{qr}
\end{equation}
Similarly for the electric current
\begin{equation}\label{eq:lincurr}
J^j=e^{jrs} G_{sq} B_{qr}
\end{equation}

Assuming incompressibility, write $\omega^j$ for $Q^j$ in \Eq{order0}
so that this represents the vorticity 
evolution~\Eq{fullpv} without the forcing terms. It follows that
the background flow is immaterial for a purely linear velocity field since then~$Q^j_i=\omega^j_i=0$.
The magnetic forcing
term in \Eq{fullpv} is also a Lie bracket hence it follows that the background magnetic field is
not dynamically significant either.

There is a problem for the linear fields' approach unless
the momentum equation is explicitly introduced, because the vorticity
equation~\Eq{linvort} represents an
evolution equation for the three components~$\omega^j$, yet there $ U^q_r$
represents $8$ or~$9$ unknowns, according as to whether the velocity field
is solenoidal or not. A good way to resolve this is to assume that
$U^q_r$ is determined by a skew-symmetric matrix, 
\begin{equation}\label{eq:skewv}
V_{qr} = e_{qrk} \varpi_k
\end{equation}
for an arbitrary vector with components~$\varpi_k$. If $\ell_{ij}$ is the transformation
matrix of coordinates, then identifying $U^q_r$ with matrix entries~$U_{qr}$,
the definition is completed as
\begin{equation}\label{eq:tfmv}
U= \ell V \ell^{-1}
\end{equation}
assuming that the transformation matrix is nonsingular
(and thus $\ell \ell^T=I$ the identity).

At this juncture, note that if analogously $B=\ell K \ell^{-1}$, then
the matrix commutator
\begin{equation}\label{eq:tfmcomm}
[U,B]= \ell [V,K] \ell^{-1}
\end{equation}
thus if $Q=B$ evolves as the commutator~\Eq{commutator}, then $\dot{K}=[V,K]$
(and vice versa). This is of course expected from the aforementioned coordinate
invariance properties of the Lie-Taylor expansion.

\subsection{Finite Domain Considerations}\label{sec:finite}
The difficulty in physically interpreting the outcome of a ``linear fields" model
is that because the fields increase linearly in an unbounded domain, they
have infinite energy at all times. (Beware that this also
implies the Helmholtz decomposition of each field into gradient and curl
is not unique.) Thus it is not unexpected that linear fields'
solutions in general have finite time singularities~\cite{Im67TwoD},
particularly when the linear fields represent inflow boundary conditions.
Imshennik and Syrovatskii~\cite{Im67TwoD} interpret these singularities as implying
current sheet formation, and go on to discuss how they might occur in
a more realistic situation where the fields are linear only in a bounded
region, inferring that the singularities require input of significant 
external energy. Hence singular linear fields cannot be regarded as
self-consistent \emph{local} models, leading to the emphasis of the
current work on ensuring bounded solutions. However, as discussed
in the introduction, there are physical difficulties regarding
the assumption of the existence of current blobs needed to ensure 
a bounded MHD problem in general.

The value of the Taylor series approach is that the higher order terms
allow for a more realistic current distribution. The question is to what
extent is their presence consistent with the simple ``linear fields" model.

\begin{figure}
\centerline{\rotatebox{0}{\includegraphics[width=0.5\columnwidth]{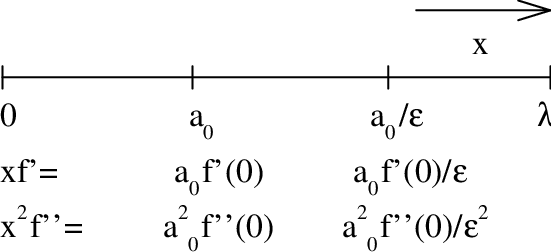}}}
\caption{Schematic of domain $0<x<\lambda$, showing origin where there is a
typically a stagnation point, and $x=a_0\ll a_0/\epsilon$, explained in the text.
$r_2=0.25$. \label{fig:line2}}
\end{figure}

The first point to notice is that the equations for evolution of the
hierarchy of derivatives in \ssec{liederiv} of \Sec{deriv} may be formally, consistently
ordered if the second and higher order derivatives are supposed to be smaller than
first order derivatives by a factor of order~$\epsilon\ll1$. Inspection of \Eq{utay}
shows that the second order term rises up to equal the first when $\epsilon r =\mathcal{O}(1)$
where $r=\|x\|/a_0$ measures distance from the origin scaled by lengthscale~$a_0$.
Writing $\lambda=a_0/\epsilon$, $\lambda$ is the lengthscale over which the quadratic terms begin to
equal the linear terms, and so, as pointed out by Arter~\cite{Wa87d},
about or beyond a distance~$\lambda$, more physically realistic boundary
conditions ensuring bounded problem energy might be imposed, see \Fig{line2}.
At $x\approx a_0$, inflow or indeed outflow boundary conditions with $u\propto \pm x$ could be imposed.

The simplest example of a function~$f$ satisfying $f''=\mathcal{O}(\epsilon) f'$ 
(prime denotes spatial derivative) is unfortunate,
namely the exponential~$\exp(x/\lambda)$, because the
rapid growth of the function with distance implies the linear
region may be hard to observe in either numerical or laboratory
experiments. Nonetheless, there will be other, less
rapidly spatially increasing functions, and the linear fields' model should be valid 
provided $\|x\|/a_0 \leq \mathcal{O}(1)$. Evidently the lengthscale~$a_0$ needs to be
smaller than the domain size~$\lambda$, and in practice it will be set by the initial ratio
of function to first derivative $\|Q^j\|/\|Q^j_i\|$ if $B^j\neq0$.

There remains the question as to whether the ordering  remains consistent
under time evolution, that is to say whether \Eq{order2} implies that second order field 
derivatives also do not grow. In the case where $U^i=0$
contracting $Q^j_{kl}$ with $\dot{Q}^j_{kl}$ from \Eq{order2} gives
\begin{equation}\label{eq:ordsq2}
Q^j_{kl}\dot{Q}^j_{kl}=
Q^j_{kl} U^i_{l}Q^j_{ik}
-Q^j_{kl} U^i_kQ^j_{il}
-Q^j_{kl} U^j_iQ^i_{kl}
\end{equation}
provided $U^j_{ki}=0$ is also assumed. In each of the three terms, the antisymmetric tensor
$U$ is contracted with a symmetric tensor consisting of the product of $Q$ with itself,
hence each vanishes separately. This statement does not depend on the coordinate system used
hence taking $Q=B$, the sum of the squares of the $B^i_{jk}$ is shown to be conserved for flows
of type \Eq{tfmv} also (the Lie-Taylor series expansion of \ssec{liederiv} of \Sec{deriv} could have been 
developed in the coordinates where $V$ was antisymmetric). Given the same result at
the end of \ssec{liederiv} of \Sec{deriv} for first order $Q$~derivatives, it should be evident
that a similar analysis could be conducted at any higher order.

\section{Case study: Dolzhansky-Kirchhoff Equations}\label{sec:case}
\subsection{Derivation}\label{sec:DZ}
The Kirchhoff equations for \emph{incompressible} MHD after Dolzhansky~\cite{Do05mech} follow 
upon assuming that the transformation of coordinates applied to an antisymmetric
matrix representation of the velocity gradient matrix~$U$ is a simple
anisotropic scaling 
\begin{equation}\label{eq:dzscale}
x_i= a_{(i)} x^i
\end{equation}
so that $\ell_{ij}$ introduced in \ssec{VandC} of \Sec{deriv} is diagonal. The convention is adopted that
suffices on~$x$, as distinct from on fields $u^i$ and~$\varrho$,  denote Cartesian coordinates.

Without the anisotropy induced by this scaling, it helps to note that Dolzhansky
has introduced a basis for the fields which takes the form 
\begin{equation}\label{eq:dzbasis}
{\bf e}_i \propto \hat{\bf x}_i \times {\bf x}
\end{equation}
where $\hat{\bf x}_1, \hat{\bf x}_2, \hat{\bf x}_3$ are the unit vectors
in Cartesian coordinates. Hence for each~$i$,
${\bf e}_i\cdot {\bf x}=0$ and the linear fields' flow expressed in terms of
this basis, viz.
\begin{equation}\label{eq:uondzb}
{\bf u}({\bf x},t) = \Sigma_{k=1}^3 \varpi"_k {\bf e}_k,
\end{equation}
where the coefficients~$\varpi"_k$ vary only with time,
has streamlines and indeed streaklines which are confined to spherical
surfaces. Under the transformation~\Eq{dzscale}, and indeed any
affine scaling, spheres become ellipsoids and the basis becomes
non-orthogonal, but the same local confinement property applies.
Rather less satisfactorily from the physical point-of-view as
mentioned in \ssec{intro}, although justification was attempted  in \ssec{finite} of \Sec{deriv}, the
magnetic field has to have the same basis and therefore is forced
to have this local confinement property, viz.
\begin{equation}\label{eq:bondzb}
{\bf B}({\bf x},t) =  \iota"_k(t) {\bf e}_k,
\end{equation}


The derivation of the Kirchhoff equations from the Lie-Taylor series approach of \Sec{deriv} proceeds by
introducing the transformed vector $\Omegab= \ell^{-1}\omegab$, in
components $\Omega^i= \ell_{ji}\omega^j$, so that \Eq{order0} for
$Q^i$ as the vorticity (i.e. with $Q^i = \omega^i$) becomes
\begin{equation}\label{eq:tfmOmeg}
\dot{\Omegab} = V \Omegab
\end{equation}
The Kirchhoff `vorticity' equation is completed
when $\Omega_i$ is expressed in terms of the entries in $\varpi_k$ in~$V$.
\Eq{linvort} and \Eq{skewv} when combined give
\begin{equation}\label{eq:omegid}
\Omega^i=e_{ijk} G_{kl} V_{lj}=e_{ijk} G_{kl} e_{ljm} \varpi_m
\end{equation}
which using Einstein's identity becomes
\begin{equation}\label{eq:omegidexp}
\Omega^i=G_{ki} \varpi_k - G_{kk} \varpi_i
\end{equation}
Taking $\ell_{ij}$ or equivalently~$G_{ij}$ to be diagonal, \Eq{omegidexp} 
yields the relation
\begin{equation}\label{eq:invomegid}
\Omega^i= I_{(i)} \varpi_i/\sqrt{g}
\end{equation}
where $I_i=g'_{ii}$ denotes e.g. that 
the term $g_{11}$ is omitted from the trace of~$g$ in the definition
of $\varpi_1$, etc.

Analogous to \Eq{skewv}, write
\begin{equation}\label{eq:skewk}
K_{qr} = e_{qrk} \iota_k
\end{equation}
and explicitly writing
\begin{equation}\label{eq:iidefn}
I_i=a_j^2+a_k^2, \;\;\; (ijk) \mbox{ a permutation of } (123)
\end{equation}
the vector vorticity equation becomes
\begin{eqnarray}\label{eq:vvort}
\dot{\varpi}_1 &=& r_1 (\varpi_2 \varpi_3 - \iota_2 \iota_3) \nonumber \\
\dot{\varpi}_2 &=& r_2 (\varpi_1 \varpi_3 - \iota_1 \iota_3) \\
\dot{\varpi}_3 &=& r_3 (\varpi_1 \varpi_2 - \iota_1 \iota_2)\nonumber
\end{eqnarray}
where 
\begin{eqnarray}\label{eq:ri}
r_1&=&(I_3-I_2)/I_1=(a^2_2-a^2_3)/(a^2_2+a^2_3) \nonumber \\
r_2&=&(I_1-I_3)/I_2=(a^2_3-a^2_1)/(a^2_1+a^2_3) \\
r_3&=&(I_2-I_1)/I_3=(a^2_1-a^2_2)/(a^2_1+a^2_2) \nonumber
\end{eqnarray}
Note that $|r_i|\leq1$ each~$i$, and that the $r_i$ are not
independent, but related by
\begin{equation}\label{eq:rrr}
r_1+r_2+r_3+r_1 r_2 r_3=0
\end{equation}
so that e.g. $r_3=-(r_1+r_2)/(1+r_1 r_2)$, and the inverse relations are for example
\begin{eqnarray}\label{eq:ir}
\frac{I_2}{I_1}&=&\frac{1-r_1}{1+r_2} \nonumber \\
\frac{I_3}{I_1}&=&\frac{1+r_1}{1-r_3}  \\
\frac{I_3}{I_2}&=&\frac{1-r_2}{1+r_3} \nonumber
\end{eqnarray}
The vector electric current equation, using \Eq{tfmcomm} is simply
\begin{eqnarray}\label{eq:kinduc}
\dot{\iota}_1 &=& \varpi_2 \iota_3 - \varpi_3 \iota_2 \nonumber \\
\dot{\iota}_2 &=& \varpi_3 \iota_1 - \varpi_1 \iota_3 \\
\dot{\iota}_3 &=& \varpi_1 \iota_2 - \varpi_2 \iota_1 \nonumber
\end{eqnarray}
Note that, as might have been anticipated, $\varpi"$ and $\varpi$, $\iota"$ and $\iota$
may respectively be identified.
Further, although the above approach may seem to enable
the generalisation of Dolzhansky's approach to a non-orthogonal coordinate
system, this does not in fact constitute a different physical situation
because the affine transformation of an ellipsoid is another
ellipsoid.

The Kirchhoff equations have a conserved Hamiltonian
\begin{equation}\label{eq:kham}
H_0=I_1 (\varpi^2_1 + \iota^2_1) + I_2 (\varpi^2_2 + \iota^2_2) + I_3 (\varpi^2_3 + \iota^2_3)
\end{equation}
together with a cross-helicity
\begin{equation}\label{eq:kxhel}
H_1=I_1 (\varpi_1 \iota_1) + I_2 (\varpi_2 \iota_2) + I_3 (\varpi_3 \iota_3)
\end{equation}
and \Eq{kinduc} as discussed in \ssec{liederiv} of \Sec{deriv} has a (Casimir) 
invariant~$C_0=\iota^2_1+\iota^2_2+\iota^2_3$.
The Kirchhoff equations may be transformed into the equations of motion
of a charged particle on a sphere, see~\cite[\S\,1]{arnoldnovikov},
although the transformation is not well explained, and not mentioned in~\cite{arnoldkhesin}.
However, the most efficient way to proceed is to note that the
Dolzhansky variant of the Kirchhoff equations is the Clebsch case, see ref~\cite[\S\,2.4]{Ho98Dyna}
and so is completely integrable with $3$~invariants 
\begin{eqnarray}\label{eq:Ci}
C_1&=&\iota^2_1+\varpi^2_2/r_2-\varpi^2_3/r_3 \nonumber\\
C_2&=&\iota^2_2-\varpi^2_1/r_1+\varpi^2_3/r_3  \\
C_3&=&\iota^2_3+\varpi^2_1/r_1-\varpi^2_2/r_2 \nonumber
\end{eqnarray}
In fact there are two remaining degrees of freedom, since $H_0=I_1C_1+I_2C_2+I_3C_3$
and $C_0=C_1+C_2+C_3$.

\subsection{Catastrophic Behaviour}\label{sec:DZcat}
It is of interest for comparison with ideal MHD to understand
the transient behaviour of Dolzhansky's
equations. Supposing that the magnetic field has negligible
dynamical effect, then it evolves kinematically in a flow,
described by vorticity variables which obey Euler's equations for the motion
of a massless spinning top in classical mechanics.
For such a body, it is a classical result that if the~$I_i$,
regarded now as moments of inertia of the top, satisfy
$I_3>I_1>I_2$ then motion about the $1$-axis is unstable.
Thus a significant transient is expected when a
slow variation of~$a_i$ is arranged such that $I_3$ approaches then drops
below~$I_1$, so that rotation about the $3$-axis is destabilised.

\begin{figure}
\centerline{\rotatebox{0}{\includegraphics[width=0.5\columnwidth]{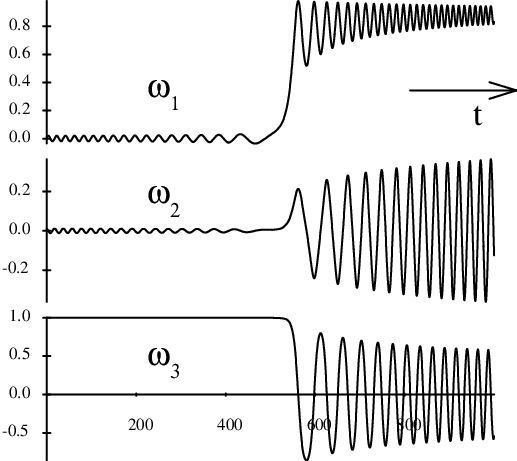}}}
\caption{Euler's equations, initial conditions $\omega_1=\omega_2=0.01$, $\omega_3=1$
$I_1=2.25$, $I_2=1.25$, $I_3(0)=2.5$, slowly varying $I_3$ satisfies
$I_3=I_3(0)-0.001 t/2$.
\label{fig:flopl}}
\end{figure}

This transient corresponds to the near disappearance (see Appendix~6(b)) 
of the effective potential
when $r_2=0$ so that the system moves ballistically on the timescale
of~$\omega_3\approx 1$ away from a now unstable equilibrium,
i.e. the rate at which the instability threshold
is crossed is of little importance. This is illustrated
in \Fig{flopl}, where the initial conditions ensure in fact that
$\iota_i=0$ for all time.
This behaviour may also be deduced from the properties of Jacobi elliptic functions
under parameter variation.

Energy remains bounded in the case of top dynamics because the
system ends in a different well with different rotation and translation direction.
The implications for field kinematics is the possibility of a sudden transient
which redirects not only the vortex but also the electric current direction.
This is reminiscent of the ideal MHD kink where components of field
and current in new directions appear. It has however to be established
that this behaviour extends to the case of a dynamically active magnetic
field.

\begin{figure}
\centerline{\rotatebox{0}{\includegraphics[width=0.5\columnwidth]{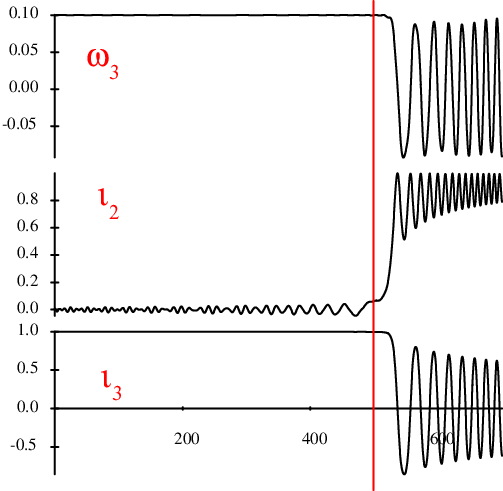}}}
\caption{Dolzhansky-Kirchhoff equations with slowly varying $r_1=-0.5+0.001 t$,
$r_2=0.25$. \label{fig:dz3ln}}
\end{figure}

\begin{figure}
\centerline{\rotatebox{0}{\includegraphics[width=0.5\columnwidth]{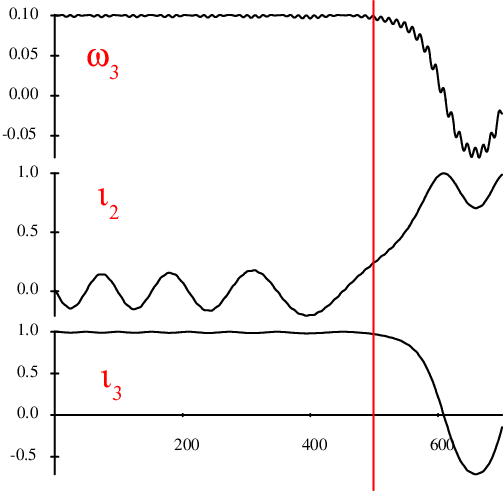}}}
\caption{Dolzhansky-Kirchhoff equations with very slowly varying $r_1=-0.005+0.000\,01 t$,
$r_2=0.25$. \label{fig:dz5ln}}
\end{figure}

\Figs{dz3ln}{dz5ln} support this contention. All the field components now have
initial conditions~$0.01$, except $\omega_3=0.1$ and $\iota_3=1$,
so that this is a study of the full Dolzhansky equations where
the magnetic field is dominant. \Fig{dz3ln} exhibits a similar
transient to the `spinning top' or field-free case when
the stability boundary $r_1=0$ (see Appendix~6(a)) 
is again crossed at the rate~$0.001$.
The exact nature of this transient requires further examination,
but it will be seen from \Fig{dz5ln} that the timescale for say
$\iota_3$ to reverse changes only by a factor of three (from
approximately $40$ to~$110$) when the crossing rate drops
by a factor of~$100$.

\subsection{Lie Algebra}\label{sec:DZlie}
The scaling \Eq{dzscale} applied to the basis functions \Eq{dzbasis} leads to
Dolzhansky's basis functions~${\bf e}_i$ (${\bf W}_i$ in the notation of ref~\cite{Do05mech})
\begin{equation}\label{eq:dzbfn}
{\bf e}_i= -\frac{a_j}{a_k}x_k \hat{\bf x}_j+\frac{a_k}{a_j}x_j \hat{\bf x}_k, \;\;\;
(ijk) \mbox{ a permutation of } (123)
\end{equation}
where $\hat{\bf x}_j$ are the usual Cartesian basis vectors, and 
the summation convention is \emph{not} employed.
These vectors are solenoidal $\nabla . {\bf e}_i=0$, hence the Lie derivative
of one basis vector with another may be efficiently evaluated as
\begin{equation}
[ {\bf e}_i, {\bf e}_j ]= \nabla \times ( {\bf e}_i \times {\bf e}_j )
\end{equation}
Direct substitution of \Eq{dzbfn}, shows that
\begin{equation}\label{eq:dzliealg}
[ {\bf e}_i, {\bf e}_j ] =  {\bf e}_k, \;\;\;
(ijk) \mbox{ even permutation of } (123)
\end{equation}
This is a particularly simple definition of a Lie algebra, since in general the
Lie bracket is allowed to be an arbitrary linear combination of
the basis vectors with coefficients known as structure constants.
The importance of \Eq{dzliealg} is that the current evolution equations \Eq{kinduc}
follow immediately from the corresponding PDE,~\Eq{induc}. However, the vorticity evolution equation involves
the Lie bracket of ${\bf e}_i$ and its curl.

The latter quantity is
\begin{equation}
\nabla \times {\bf e}_i = (\frac{a_j}{a_k} + \frac{a_k}{a_j}) \hat{\bf x}_i, \;\;\;
(ijk) \mbox{ a permutation of } (123)
\end{equation}
Choosing ${\bf c}_i = a_{(i)}\hat{\bf x}_i$, $\nabla \times {\bf e}_i= I_{(i)} {\bf c}_i/\sqrt{g}$ and
\begin{equation}
[{\bf e}_i,{\bf c}_j]={\bf c}_k, \;\;\;
(ijk) \mbox{ even permutation of } (123)
\end{equation}
and obviously $[{\bf c}_i,{\bf c}_j]={\bf 0}$. Given these new Lie brackets,
the ODE,~\Eq{vvort}, may be written down immediately from the PDE,~\Eq{fullpv}.

\section{Resistivity}\label{sec:resist}
Resistivity, either in the classical isotropic case, or possibly as the
result of a renormalisation approach to turbulence, leads to an 
additional term in the induction equation, which assuming $\rho=\mbox{const.}$, 
becomes
\begin{equation}
\frac{\partial {\bf B}}{\partial t}=[{\bf u},{\bf B}]+\nabla\times\eta\nabla\times{\bf B},
\end{equation}
If ${\bf B}$ has a linear field representation in Cartesian coordinates,
then $\nabla\times\eta\nabla\times{\bf B}= \nabla\eta\times\nabla\times{\bf B}$.
Considering first the spherical, or unscaled case as discussed at the
start of \ssec{DZ} of \Sec{case},
the resistive term contains contributions of the form~$\nabla\eta\times\hat{\bf x}_i$.
These new terms are simply expressible in Dolzhansky's
basis ${\bf e}_i$ if $\nabla\eta \propto {\bf x}$.
It is then plausible, as may be verified by direct substitution,
that in scaled coordinates, if the resistivity is written
\begin{equation}\label{eq:quadeta}
\eta = \eta_0 +\boldsymbol{\eta}_1\cdot{\bf x} + \eta_2\left(
\frac{x^2_1}{a^2_1}+\frac{x^2_2}{a^2_2}+\frac{x^2_3}{a^2_3} \right)
\end{equation}
then this simple relationship still applies
($\eta_0$ and $\eta_1$ do not directly affect the model evolution
of linear field~${\bf B}$).

It follows that
for quadratic spatial dependence of resistivity of the form \Eq{quadeta},
the  equations for electric current variation acquire terms
\begin{equation}
\eta_2 \left(\frac{1}{a_j^2}+\frac{1}{a_k^2}\right)\iota_i,\;\;\;
(ijk)  \mbox{ permutation of } (123)
\end{equation}
and become
\begin{eqnarray}\label{eq:resinduc}
\dot{\iota}_1 &=& \varpi_2 \iota_3 - \varpi_3 \iota_2+ 
\eta_2 \left(\frac{1}{a_2^2}+\frac{1}{a_3^2}\right)\iota_1 \nonumber\\
\dot{\iota}_2 &=& \varpi_3 \iota_1 - \varpi_1 \iota_3+
\eta_2 \left(\frac{1}{a_1^2}+\frac{1}{a_3^2}\right)\iota_2 \\
\dot{\iota}_3 &=& \varpi_1 \iota_2 - \varpi_2 \iota_1 +
\eta_2 \left(\frac{1}{a_1^2}+\frac{1}{a_2^2}\right)\iota_3 \nonumber
\end{eqnarray}
Note that $\eta$ must be maximal at the origin ($\eta_2<0$) otherwise
the new resistivity terms imply exponential growth of the square
current~$(\iota^2_1+\iota^2_2+\iota^2_3)$.
Although this conclusion seems rather strange, it is in fact
consistent with the conclusion drawn by 
Forbes at al~\cite{Fo13Inde}, who find that only for locally
maximum resistivity is the Petschek mechanism for reconnection
structurally stable.

\section{Conclusion}\label{sec:conc}
This work has reinforced the contentions of ref~\cite{Do05mech}
that the Dolzhansky-Kirchhoff (D-K) equations
exhibit mathematical properties important for understanding
nonlinear magnetohydrodynamics (MHD) in the limit of small or vanishing resistivity.
\Sec{deriv} illustrates the general mathematical framework into which
the equations fit, then \ssec{DZ} of \Sec{case} shows how the D-K equations emerge when a
solution for ideal MHD is sought as a Taylor series in Cartesian
coordinates. \ssec{DZcat} of \Sec{case} shows that catastrophism is natural in the system,
in a consistent sense, namely that model variables remain bounded,
despite the dynamic timescale.

The derivation of the model also illustrates important features
of the Lie derivative, specifically its anti- or skew-symmetric
nature as demonstrated by its replacement by the matrix-commutator
in \Sec{deriv}.
In \Sec{deriv} also, important conservation relations, extending to arbitrary
order of power of the unknowns, are illustrated. Continuing the mathematical note,
\ssec{DZlie} of \Sec{case} shows the importance of the concept of Lie algebra when
seeking time-dependent solutions of nonlinear MHD.
All these properties imply that the D-K equations should also be
an aid to understanding the reduction of PDEs to non-canonical
Hamiltonian systems and subsequent analysis~\cite{Ho98Eule,arnoldkhesin}.


As already incidentally demonstrated by the catastrophic
simulations, the sixth order D-K equations, with four invariants, are capable
of exhibiting oscillation with two different timescales and amplitudes 
over a wide range of frequencies. The reduction of nonlinear MHD to
a low order Hamiltonian system highlights the likely prevalence of
oscillation in ideal MHD, since such Hamiltonian systems do not generically
possess attracting steady solutions and their stability can only be established
in a Lyapunov sense. It is plausible that these considerations extend
to the case of small resistivity when current sheets do not form.

Other important physical behaviour, such as the existence and
behaviour of nonlinear Alfvenic solutions, corresponding to $\iota_k = c_{(k)} \varpi_k$,
for some constants~$c_k$, may be deduced 
from known results for the Kirchhoff equations~\cite{Ho98Dyna}.

\Sec{resist} shows how resistivity may be included in the model
so that the physics of reconnection, believed important in many
laboratory and astrophysical contexts, may be studied.
In the context of laboratory plasmas, specifically the central
region of tokamaks, the structural stability question raised by Forbes~\emph{et al} requires
further analysis.

Dolzhansky~\cite{Do05mech}
describes how other important physics such as rotation and gravity
(with buoyancy force thanks to a
density evolving due to a thermal evolution equation), may be
included. Another application of linear fields, which will require development
of the Lie-Taylor expansion
for the momentum conservation equation along the lines of the
mass conservation equation in \ssec{scalar} of \Sec{deriv}, is to partially ionised plasma where mass and momentum sources
allow \emph{out}flow boundary conditions, e.g. in 1-D models of the tokamak edge.

Further mathematical and physical insights into these more
complicated inviscid or almost-inviscid situations may be anticipated.
As touched upon in ref~\cite{Wa13c}, the vorticity evolution equation presents the
problem that the algebra must involve not only basis functions
but their curls. Progress may be made using Beltrami or `screw' fields since
these offer the potential to explore nonorthogonal geometries without
the explicit appearance of the metric tensor, but
this will be discussed elsewhere.

\competing{I have no competing interests.}
\ack{
The ODE integrations were performed using Hindmarsh's LSODE package.}
\funding{
This work has been funded by the RCUK Energy Programme [grant number EP/I501045].}
\dataccess{ To obtain further information on the data and models underlying this
paper please contact PublicationsManager@ccfe.ac.uk.}

\section{Appendices}\label{sec:app}
\subsection{Stability Analysis}\label{sec:stab}
Note that $\omega$ is used as a synonym for $\varpi$ in the following.
\subsubsection{Aligned Vorticity and Current}\label{sec:align}
Without loss of generality, assume that the vorticity and current are aligned in
the direction of the $3$-axis, with $\omega_3=W$ and $\iota_3=J$ and all
other components of~$\mathcal{O}(\epsilon)$ at time $t=0$. With these
assumptions \Eq{vvort} and \Eq{kinduc} become respectively
\begin{eqnarray}\label{eq:vvl}
\dot{\omega}_1 &=& r_1 (W\omega_2 - J \iota_2) \\
\dot{\omega}_2 &=& r_2 (W \omega_1 - J \iota_1) \label{eq:vvl2}\\
\dot{\omega}_3 &=& \mathcal{O}(\epsilon^2) \nonumber
\end{eqnarray}
\begin{eqnarray}\label{eq:kil}
\dot{\iota}_1 &=& J \omega_2  - W \iota_2  \\
\dot{\iota}_2 &=& W \iota_1 - J \omega_1 \label{eq:kil2}\\
\dot{\iota}_3 &=& \mathcal{O}(\epsilon^2) \nonumber
\end{eqnarray}
Differentiating \Eq{vvl2} with respect to time, and substituting for first derivatives
using \Eq{vvl} and \Eq{kil} gives
\begin{equation}\label{eq:w2evol}
\ddot{\omega}_2= r_2 \omega_2 (r_1 W^2-J^2)+r_2J W (1-r_1)\iota_2
\end{equation}
Similarly differentiating \Eq{kil2} with respect to time, and substituting for first derivatives
using \Eq{vvl} and \Eq{kil} gives
\begin{equation}\label{eq:j2evol}
\ddot{\iota}_2= (r_1 J^2-W^2)\iota_2 +J W (1-r_1)\omega_2
\end{equation}

Seeking solutions varying in time $\propto \exp(s t)$ to \Eqs{w2evol}{j2evol} gives the
determinantal equation $\det(M)=0$ where
\begin{equation}
M=\begin{pmatrix}
-s^2+m_{11} & m_{12} \\
r_2 m_{12} & -s^2+m_{22}
\end{pmatrix}
\end{equation}
with $m_{11}=r_1 J^2-W^2$, $m_{12}=JW(1-r_1)$ and $m_{22}=r_2(r_1 W^2-J^2)$.
The resulting stability polynomial is 
\begin{equation}\label{eq:stabpoly}
s^4 -\left(J^2(r_1-r_2)+W^2(r_1r_2-1)\right) s^2 -r_1(J^2-W^2)^2=0
\end{equation}
Instability may be avoided if both roots of the corresponding quadratic (with $x=s^2$)
are negative, implying $\beta<0$ and $r_1<0$, where $\beta$ is the coefficient of~$s^2$.
The former means
\begin{equation}\label{eq:r2ineq}
r_2> \frac{(r_1 J^2-W^2)}{(J^2-r_1W^2)}
\end{equation}
Note that this stability analysis may be checked by differentiating \Eqs{vvl}{kil} with respect to time
and substituting using \Eqs{vvl2}{kil2}. Numerical solution in the text confirms
that $r_1=0$ is indeed a stability boundary.

\subsubsection{Orthogonal Vorticity and Current}\label{sec:nonalign}
Without loss of generality, assume that the vorticity and current are aligned in
the directions of the $3$-axis and the $1$-axis respectively,
with $\omega_3=X$ and $\iota_1=K$ and all
other components of~$\mathcal{O}(\epsilon)$ at time $t=0$. With these
assumptions \Eq{vvort} and \Eq{kinduc} become respectively
\begin{eqnarray}\label{eq:vwl}
\dot{\omega}_1 &=& r_1 X \omega_2 \\
\dot{\omega}_2 &=& r_2 (X \omega_1 - K \iota_3) \label{eq:vwl2}\\
\dot{\omega}_3 &=& -r_3 K \iota_2 \nonumber
\end{eqnarray}
\begin{eqnarray}\label{eq:kjl}
\dot{\iota}_1 &=& - X \iota_2  \\
\dot{\iota}_2 &=& \iota_1  \omega_3 = XK \label{eq:kjl2}\\
\dot{\iota}_3 &=& -K \omega_2 \nonumber
\end{eqnarray}
\Eq{kjl2} shows that $\iota_2$ grows on a $\mathcal{O}(1)$ timescale unless
either $X=o(1)$ or $K=o(1)$.

In the former case $X=o(1)$, differentiating \Eq{vwl2}  gives
\begin{equation}\label{eq:w2dot}
\ddot{\omega}_2 = r_2 K^2 \omega_2 
\end{equation}
so there might be stability if $r_2<0$,
but even so $\omega_3$ varies in time proportional to~$\iota_2$.
Similarly in the latter case $K=o(1)$
\begin{equation}\label{eq:w2dotl}
\ddot{\omega}_2 = r_1 r_2 X^2 \omega_2 
\end{equation}
and there is again an apparently secular variation, this time in~$\iota_1$.

Thus it seems that there is no stable steady solution with orthogonal
vorticity and current. This conclusion is supported by analysis
with $K,\;X=\mathcal{O}(1)$ that follows if \Eq{kjl2} is differentiated with respect to time
without assuming that $\omega_3$ and~$\iota_1$ are constant. For then
\begin{equation}\label{eq:w3dot}
\ddot{\omega}_3 = -r_3 K^2 \omega_3 
\end{equation}
implying that $\omega_3$ oscillates with frequency $\sqrt{r_3}K$ (provided $r_3>0$).
It also follows that $\iota_1$ and  $\iota_2$ oscillate with frequencies~$X$
and $\sqrt{(X^2+r_3K^2)}$ respectively provided the $\omega_3$ oscillation is slow.

For the $\omega_1$ and~$\iota_3$ dynamic, there is a 
determinantal equation $\det(M')=0$ where
\begin{equation}
M'=\begin{pmatrix}
-s^2+m'_{11} & r_1 m'_{12} \\
m'_{12} & -s^2+m'_{22}
\end{pmatrix}
\end{equation}
with $m'_{11}=r_1 r_2X^2$, $m'_{12}=r_2KX$ and $m'_{22}=r_2K^2$. This leads to
a quadratic in $x=s^2$ with roots $x=0$ and
\begin{equation}\label{eq:stabx}
x =r_2 (K^2+r_1X^2)
\end{equation}
This may imply instability unless $r_2<0$ and $r_1 < -K^2/X^2$ or
$r_2>0$ and $r_1 > -K^2/X^2$.

\subsection{`Disappearance' of the Potential}\label{sec:disapp}
The surprising, at first hearing, remark that the potential almost disappears when $I_3=I_1$,
is explained if each component of vorticity is treated as evolving in its own
separate potential. The relevant equations may be derived by differentiating
each of the equations for $\varpi_i=\omega_i$ in \Eq{vvort} separately with respect to time, then
eliminating first derivatives in terms of the $\omega_i$.
This gives equations specifiable, given
\begin{equation}\label{eq:ddw3}
\ddot{\omega}_3 = r_3 \omega_3 (r_1 \omega_2^2+ r_2 \omega_1^2),
\end{equation}
as even cyclic permutations of the equation's suffices~$(123)$.
Substituting for $C_1,\;C_2,\;C_3$ in \Eq{ddw3} and permutations then gives
\begin{eqnarray}
\nonumber
\ddot{\omega}_1 &=& r_2 r_3 \omega_1 \left( (C_2-C_3)r_1 + 2 \omega_1^2\right) \\
\ddot{\omega}_2 &=& r_1 r_3 \omega_2 \left( (C_3-C_1)r_2 + 2 \omega_2^2\right) \label{eq:ddw}\\
\ddot{\omega}_3 &=& r_1 r_2 \omega_3 \left( (C_1-C_2)r_3 + 2 \omega_3^2\right) 
\nonumber
\end{eqnarray}

Each equation~\Eq{ddw} represents the motion of a particle in a potential~$V_i$, e.g. given by
\begin{equation}\label{eq:w3v}
2 V_3(\omega_3)=r_1 r_2  (C_1-C_2)r_3 \omega^2_3+ r_1 r_2  \omega^4_3
\end{equation}
in the case of~$\omega_3$, and obvious permutations for the other~$\omega_i$.
When $I_3=I_1$, $r_2=0$ from \Eq{ri}, and it follows that the quadratic
potential contributions to $V_i$ vanish for each~$i$, and only the quartic term
in $V_2$ survives. The corresponding variable is~$\omega_2$ which is small at least
initially, hence the particle representing vector~$\omegab$ is able to move ballistically,
as though there were no potential forces present.

The table gives values to $2$~decimal places for the coefficients of $2V_i$
at the start of the simulation shown in \Fig{flopl}, all of which change
sign at $r_2=0$ except the one marked with a dagger. The corresponding dependence 
of the potential~$V_i$ on~$\omega_i$
is denoted by `U',`W' or `M', with the letter corresponding to the approximate shape of
the potential. Thus $\omega_3$ sits stably in the well at the bottom of the
right-hand of the `W'~potential, which flips to become an 'M' shape in $r_2>0$ whereupon $\omega_3$ 
oscillates in the well in the centre of the `M'.
\begin{table}\label{tab:pots}
\caption{Coefficients of effective potentials for Euler's equations
at start of simulation.}
\begin{center}
\begin{tabular}{c|ccc}
Term & $2V_1$ & $2V_2$ & $2V_3$ \\  
\hline
$\omega^2$ & $-0.11$ & $-0.12$  & $0.24$ \\  
$\omega^4$ & $0.08$ & $-0.24$\dag  & $-0.12$ \\  
\hline
$r_2<0\mapsto >0$& $M\mapsto W$ & $U\mapsto W$  & $W\mapsto M$ \\
\hline
& \dag does not change sign.
\end{tabular}
\end{center}
\end{table}


\end{document}